\RequirePackage{lineno}
\documentclass[aps,prd,twocolumn,letterpaper,superscriptaddress, showpacs, amsmath, amssymb]{revtex4}

\usepackage{graphicx}
\usepackage{amsmath,amssymb}
\usepackage{color}

\begin{document}


\title{Testing the hadronuclear origin of PeV neutrinos observed with IceCube}

\author{Kohta Murase}
\affiliation{Hubble Fellow -- Institute for Advanced Study, Princeton, New Jersey 08540, USA}
\author{Markus Ahlers}
\affiliation{Wisconsin IceCube Particle Astrophysics Center (WIPAC) and Department of Physics, University of Wisconsin, Madison, Winsconsin 53706, USA}
\author{Brian C. Lacki}
\affiliation{Jansky Fellow of the National Radio Astronomy Observatory -- Institute for Advanced Study, Princeton, New Jersey 08540, USA}

\date{submitted 15 June 2013; published 3 December 2013}

\begin{abstract}
We consider implications of the IceCube signal for hadronuclear ($pp$) scenarios of neutrino sources such as galaxy clusters/groups and star-forming galaxies.  Since the observed neutrino flux is comparable to the diffuse $\gamma$-ray background flux obtained by {\it Fermi}, we place new, strong {\it upper} limits on the source spectral index, $\Gamma\lesssim2.1\mbox{--}2.2$.  In addition, the new IceCube data imply that these sources contribute {\it at least} $30$\%$\mbox{--}40$\% of the diffuse $\gamma$-ray background in the 100~GeV range and even $\sim100$\% for softer spectra.  Our results, which are insensitive to details of the $pp$ source models, are one of the first strong examples of the multimessenger approach combining the {\it measured} neutrino and $\gamma$-ray fluxes.  The $pp$ origin of the IceCube signal can further be tested by constraining $\Gamma$ with sub-PeV neutrino observations, by unveiling the sub-TeV diffuse $\gamma$-ray background and by observing such $pp$ sources with TeV $\gamma$-ray detectors.  We also discuss specific $pp$ source models with a multi-PeV neutrino break/cutoff, which are consistent with the current IceCube data.
\end{abstract}

\pacs{95.85.Ry, 98.70.Sa, 98.70.Vc\vspace{-0.3cm}}
\maketitle

%
\section{Introduction}
High-energy neutrinos provide the ``smoking-gun'' signal of cosmic-ray (CR) acceleration~\cite{neurev}, and their detection with the IceCube observatory has long been anticipated~\cite{icecube}.  In 2012, the IceCube Collaboration announced the detection of two PeV shower events observed during the combined IC-79/IC-86 data period~\cite{PeV2}.  A recent follow-up analysis~\cite{PeV28} of the same data uncovered a spectrum of 26 additional events at lower energies.  These new data are consistent with an isotropic neutrino background (INB) flux of $E_{\nu}^2\Phi_{\nu_i}\sim{10}^{-8}~{\rm GeV}~{\rm cm}^{-2}~{\rm s}^{-1}~{\rm sr}^{-1}$ (per flavor) around PeV~\cite{PeV2,PeV28}, in agreement with the conventional Waxman-Bahcall bound~\cite{wbbound}.  A break/cutoff at $E_\nu\sim1\mbox{--}2$~PeV is suggested for hard spectra with spectral indices of $\Gamma\sim2$, since no events were found at higher energies where the effective area is larger especially due to the Glashow resonance at 6.3~PeV~\cite{PeV28,lah+13}.  

The origin of the IceCube signal is unknown.  Among extragalactic neutrino sources, jets and cores of active galactic nuclei (AGN)~\cite{agnjet,agncore} and $\gamma$-ray burst (GRB) jets~\cite{grb,llgrb} have been widely studied, where the photohadronic (e.g., $p\gamma$) reaction is typically the main neutrino generation process.  On the other hand,  large scale structures with intergalactic shocks (IGSs) and AGN~\cite{mur+08,gc}, and starburst galaxies (SBGs)~\cite{sbg} may significantly contribute to the INB mainly via the hadronuclear (e.g., $pp$) reaction.  
It is crucial to discriminate $pp$ and $p\gamma$ scenarios to identify the neutrino sources.  In this work, we consider the $pp$ origin and show that it can be tested with the multimessenger approach in the next several years.  

Recently, {\it Fermi} improved limits on the diffuse isotropic $\gamma$-ray background (IGB) by $\sim10$ times compared to EGRET~\cite{egb}, so the known connection between the INB and the IGB~\cite{bs75} leads to stronger constraints on neutrino emission.  Although $p\gamma$ emission, especially cosmogenic signal, has been the main interest~\citep[e.g.,][]{connect,mur+12}, $pp$ sources have not been explicitly studied.  There is an important difference between the $pp$ and $p\gamma$ cases.  In $p\gamma$ scenarios, secondary spectra typically have a strong energy dependence (rising at $\gg$~GeV energies) due to the threshold and dominance of resonant channels.  In contrast, the approximate Feynman scaling of $pp$ reactions leads to power-law secondary spectra stretching from GeV energies, following the initial CR spectrum.  Hence, normalization of the neutrino spectrum at PeV energies has immediate consequences on $\gamma$-ray spectra at lower energies, giving us powerful constraints on $pp$ scenarios.  

The new IceCube data show that the total INB flux is {\it comparable} to the diffuse IGB flux~\cite{PeV2,PeV28}.  This enables us to obtain the allowed range in viable $pp$ scenarios for the first time.  Our conclusion that the $pp$ sources {\it must} have $\Gamma\lesssim2.1\mbox{--}2.2$ implies that the $pp$ origin can be tested by (a) determining $\Gamma$ by IceCube, (b) resolving sources by {\it Fermi} and understanding the diffuse IGB, and (c) observing more individual sources with TeV $\gamma$-ray telescopes, especially the future Cherenkov Telescope Array (CTA)~\cite{cta}.  Our results are insensitive to redshift evolution and even remain valid for Galactic sources when we regard the observed neutrino flux as isotropic.  We briefly discuss specific sources with a neutrino break/cutoff around PeV.  
Throughout this work, we use $A_x=A/{10}^x$ and cosmological parameters with $h=0.71$, $\Omega_m=0.3$, and $\Omega_\Lambda=0.7$.  

\section{The Multimessenger Connection}
We generally consider $pp$ neutrinos produced inside the sources, which more specifically include galaxy clusters (GCs) and star-forming galaxies (SFGs).  For the $pp$ reaction at sufficiently high energies, the typical neutrino energy is 
\begin{equation}
E_\nu\sim0.04E_p\simeq2~{\rm PeV}~\varepsilon_{p,17}[2/(1+\bar{z})],
\end{equation}
where $\varepsilon_p={10}^{17}~{\rm eV}~\varepsilon_{p,17}$ is the proton energy in the cosmic rest frame and $\bar{z}$ is the typical source redshift.  Neutrinos around a possible $\sim1\mbox{--}2$~PeV break come from protons with energies close to the {\it second/iron knee}~\cite{mur+08}.  Note that the neutrino energy is less for nuclei with the same energy, since the energy per nucleon is lower.  The energy per nucleon should exceed the {\it knee} at $3\mbox{--}4$~PeV.  

Given the differential CR energy budget at $z=0$, $Q_{E_p}$, the INB flux per flavor is estimated to be~\cite{wbbound,mur+08} 
\begin{eqnarray}
E_{\nu}^2\Phi_{\nu_i}\approx\frac{ct_H\xi_z}{4\pi}\frac{1}{6}{\rm min}[1,f_{pp}](E_pQ_{E_p})
\end{eqnarray}
where $t_H\simeq13.2$~Gyr and $\xi_z$ is the redshift evolution factor~\cite{wbbound,mur+12}.  The $pp$ efficiency is
\begin{equation}
f_{pp}\approx n\kappa_{p}\sigma_{pp}^{\rm inel}ct_{\rm int},
\end{equation}
where $\kappa_{p}\approx0.5$, $\sigma_{pp}^{\rm inel}\sim8\times{10}^{-26}~{\rm cm}^2$ at $\sim100$~PeV~\cite{pp}, $n$ is the typical target nucleon density, $t_{\rm int}\approx{\rm min}[t_{\rm inj},t_{\rm esc}]$ is the duration that CRs interact with the target gas, $t_{\rm inj}$ is the CR injection time and $t_{\rm esc}$ is the CR escape time.  

The $pp$ sources we consider should also contribute to the IGB.  As in Eq.~(2), their generated IGB flux is
\begin{equation}
E_{\gamma}^2\Phi_{\gamma}\approx\frac{ct_H\xi_z}{4\pi}\frac{1}{3}{\rm min}[1,f_{pp}](E_pQ_{E_p}),
\end{equation}
which is related to the INB flux \textit{model independently} as 
\begin{equation}
E_{\gamma}^2\Phi_{\gamma}\approx2(E_{\nu}^2\Phi_{\nu_i})|_{E_\nu=0.5E_\gamma}.
\end{equation}
Given $E_{\nu}^2\Phi_{\nu_i}$, combing Eq.~(5) and the upper limit from the \textit{Fermi} IGB measurement $E_{\gamma}^2\Phi_{\gamma}^{\rm up}$ leads to $\Gamma\leq2+\ln[E_{\gamma}^2\Phi_\gamma^{\rm up}|_{100~{\rm GeV}}/(2E_{\nu}^2\Phi_{\nu_i}|_{E_{\nu}})]{[\ln(2 E_{\nu}/{100~{\rm GeV}})]}^{-1}$.  Using $E_{\nu}^2\Phi_{\nu_i}={10}^{-8}~{\rm GeV}~{\rm cm}^{-2}~{\rm s}^{-1}~{\rm sr}^{-1}$ as the measured INB flux at $0.3$~PeV~\cite{PeV2,PeV28,whi13}, we obtain
\begin{equation}
\Gamma\lesssim2.185\left[1+0.265\log_{10}\left(\frac{(E_{\gamma}^2\Phi_\gamma^{\rm up})|_{100~{\rm GeV}}}{{10}^{-7}~{\rm GeV}~{\rm cm}^{-2}~{\rm s}^{-1}~{\rm sr}^{-1}}\right)\right].
\end{equation}
Surprisingly, the {\it measured} (all flavor) INB flux is comparable to the {\it measured} diffuse IGB flux in the sub-TeV range, giving us new insights into the origin of the IceCube signal; source spectra of viable $pp$ scenarios must be quite hard.  Numerical results, considering intergalactic electromagnetic cascades~\cite{mb12} and the detailed {\it Fermi} data~\cite{egb}, are shown in Figs.~1-3.  We derive the strong {\it upper} limits of $\Gamma\lesssim2.1\mbox{--}2.2$, consistent with Eq.~(6).  In addition, we first obtain the {\it minimum} contribution to the 100~GeV diffuse IGB, $\gtrsim30$\%$\mbox{--}40$\%, assuming $\Gamma\geq2.0$.  Here, the IGB flux at $\sim100$~GeV is comparable to the generated $\gamma$-ray flux (see Fig.~3) since the cascade enhancement compensates the attenuation by the extragalactic background light, enhancing the usefulness of our results.  Also, interestingly, we find that $pp$ scenarios with $\Gamma\sim2.1\mbox{--}2.2$ explain the ``very-high-energy excess''~\cite{mur+12} with no redshift evolution, or the multi-GeV diffuse IGB with the star-formation history, which may imply a common origin of the INB and IGB.  

\begin{figure}[t]
\includegraphics[width=3.00in]{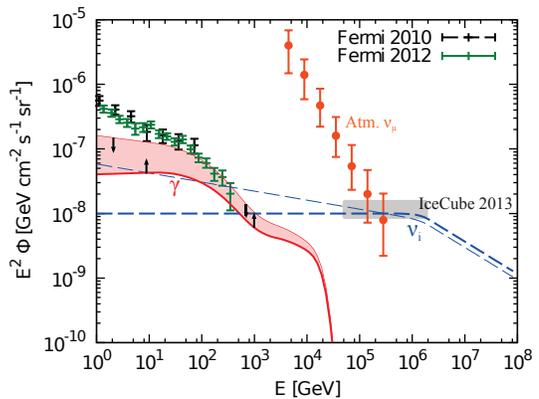}
\caption{The allowed range in $pp$ scenarios explaining the {\it measured} INB flux, which is indicated by the shaded area with arrows.  With no redshift evolution, the INB (dashed) and corresponding IGB (solid) are shown for $\Gamma=2.0$ (thick) and $\Gamma=2.14$ (thin).  The shaded rectangle indicates the IceCube data~\cite{PeV28}.  The atmospheric muon neutrino background~\cite{atm} and the diffuse IGB data by \textit{Fermi}/LAT~\cite{egb} are depicted.  
\label{fig1}
}
\vspace{-1.\baselineskip}
\end{figure}

\begin{figure}[t]
\includegraphics[width=3.00in]{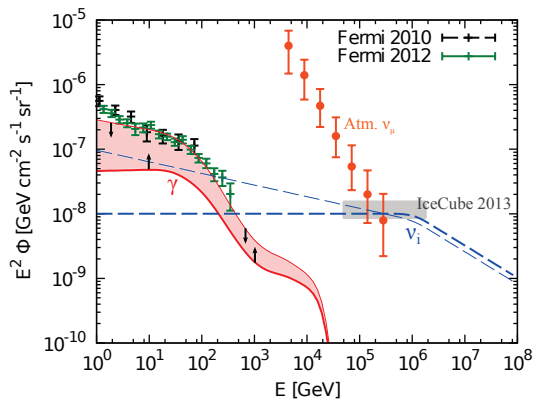}
\caption{The same as Fig.~1, but for $\Gamma=2.0$ (thick) and $\Gamma=2.18$ (thin) with the star-formation history~\cite{hb06}.  
\label{fig2}
}
\vspace{-1.\baselineskip}
\end{figure}

\begin{figure}[t]
\includegraphics[width=3.00in]{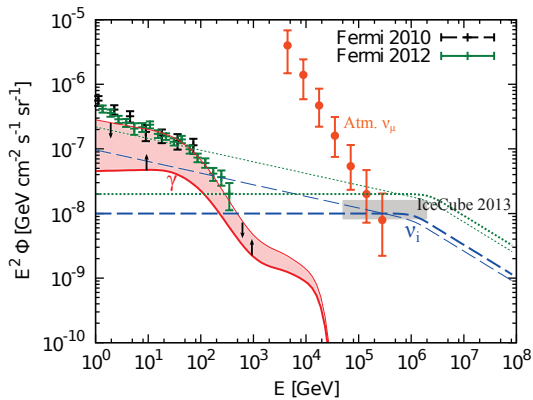}
\caption{The same as Fig. 1, but for $\Gamma=2.0$ (thick) and $\Gamma=2.18$ (thin) with the redshift evolution of $\propto{(1+z)}^3$ for $z\leq1$ and $\propto{(1+z)}^0$ for $z>1$.  The generated $\gamma$-ray spectra (dotted) before electromagnetic cascades are also shown.  
\label{fig3}
}
\vspace{-1.\baselineskip}
\end{figure}

Importantly, our results are insensitive to redshift evolution models.  In Fig.~3, we consider the different redshift evolution.  
But the result is essentially similar to those in Figs.~1 and 2.  In Figs.~1-3, the maximum redshift is set to $z_{\rm max}=5$, while we have checked that the results are practically unchanged for different $z_{\rm max}$.  This is because $\xi_z$ in Eqs.~(2) and (4) is similar and cancels out in obtaining Eq.~(5).  This conclusion largely holds even if neutrinos and $\gamma$ rays are produced at very high redshifts.  Interestingly, our results are applicable even to {\it unaccounted-for} Galactic sources, since the diffuse IGB is a residual isotropic component obtained after subtracting known components including diffuse Galactic emission.  If we use the preliminary {\it Fermi} data, based on the unattenuated $\gamma$-ray flux in Fig.~3, only $\Gamma\sim2.0$ is allowed.  

Note that such powerful constraints are not obtained for $p\gamma$ scenarios.  First, $p\gamma$ reactions are typically efficient only for sufficiently high-energy CRs, so the resulting $\gamma$ rays can contribute to the IGB only via cascades -- low-energy pionic $\gamma$ rays do not directly contribute and the differential flux is reduced by their broadband spectra, as demonstrated in~\cite{kal+13}.  More seriously, in $p\gamma$ sources like GRBs and AGN, target photons for $p\gamma$ reactions often prevent GeV-PeV $\gamma$ rays from leaving the source, so the connection is easily lost~\cite{der+07}.  Furthermore, synchrotron cooling of cascade $e^{\pm}$ may convert the energy into x rays and low-energy $\gamma$ rays, for which the diffuse IGB is not constraining.  In contrast, $pp$ sources considered here are transparent up to $\gtrsim10\mbox{--}100$~TeV energies~\cite{mb13,lt13}.  

We arrive at the following implications for $pp$ scenarios: 
(a) The spectral index should be hard, $\Gamma\lesssim2.1\mbox{--}2.2$, consistent with the present IceCube data.  However, future observations in the sub-PeV range can reasonably determine $\Gamma$ in several years~\cite{lah+13,whi13}.  For example, if $\Gamma\sim2.3$ as suggested in~\cite{anc+13}, the IceCube signal will support the $p\gamma$ origin whether the INB is Galactic or extragalactic.  
(b) The minimum contribution to the diffuse IGB is $\gtrsim30$\%$\mbox{--}40$\%.  Resolving more sources and understanding the IGB can tighten the constraints.  It is widely believed that unresolved blazars account for $\gtrsim50$\% of the diffuse IGB at $\gtrsim100$~GeV~\cite{egb,recentblazar}, which gives $\Gamma\sim2.0\mbox{--}2.1$.  If $\gtrsim60$\%$\mbox{--}70$\% of the diffuse IGB comes from them, $pp$ scenarios are disfavored.  
Better modeling of specific $pp$ sources is also useful.  For example, some predictions of $\gamma$ rays from SFGs account for $\lesssim40$\% of the diffuse IGB with $\Gamma=2.2$~\cite{lac+13,ack+12,recentsfg}, where it is difficult for the SFGs to explain the measured INB.  (c) Intrinsic $\gamma$-ray spectra of individual sources, if detected, should be hard as well.  When $pp$ sources like SFGs and IGSs significantly contribute to the diffuse IGB, as required, deeper $\gtrsim0.1$~TeV observations by, e.g., CTA will find more known $\gamma$-ray sources like SBGs or may detect sources like GCs that have not been firmly established as $\gamma$-ray sources.  These also give us crucial clues to more specific scenarios.  

In Figs.~1-3, broken power-law spectra (with $\Gamma_2=2.5$ above the break at $\varepsilon_\nu^b=2$~PeV) are used.  Importantly, our results are valid even without the break/cutoff, since they are essentially determined by $\lesssim1$~PeV emission.  

\section{Specific Scenarios}
Viable scenarios must have sufficient CR energy budget and $pp$ efficiency.  Using $E_pQ_{E_p}=Q_{\rm cr}/{\mathcal R}_p$ (where $Q_{\rm cr}\equiv\int dE_p\,\,Q_{E_p}$ is the total budget), for a power-law CR spectrum, Eq.~(2) becomes
\begin{eqnarray}
E_{\nu}^2\Phi_{\nu_i}&\simeq&1.3\times{10}^{-8}~{\rm GeV}~{\rm cm}^{-2}~{\rm s}^{-1}~{\rm sr}^{-1}~(\xi_z/3){(25/{\mathcal R}_p)}\nonumber\\
&\times&{({\rm min}[1,f_{pp}]Q_{\rm cr}/{10}^{45}~{\rm erg}~{\rm Mpc}^{-3}~{\rm yr}^{-1})},
\end{eqnarray}
where ${\mathcal R}_p\sim18\mbox{--}27$ for $s=2$ and ${\mathcal R}_p\sim200$ for $s=2.2$ (at $\varepsilon_p=100$~PeV) and $s$ is the CR spectral index.  Here, we show that large scale structures and SFGs can explain the IceCube signal~\cite{PeV2,PeV28} within uncertainty.  Note that $pp$ scenarios require $\Gamma\lesssim2.1\mbox{--}2.2$ even if the break/cutoff is absent.  On the other hand, as independently indicated in~\cite{PeV28,lah+13}, the break/cutoff is favored for such hard spectra due to significantly larger effective areas at multi-PeV energies~\cite{whi13}, so it is interesting to discuss its origin.    

\subsection{Galaxy clusters/groups}
AGN including radio galaxies are located in large scale structures containing GCs and galaxy groups.  Radio galaxies with the jet luminosity of $L_j\sim{10}^{43-47}~{\rm erg}~{\rm s}^{-1}$~\cite{agnpower} are promising CR accelerators, leading to the CR budget of $Q_{\rm cr}\sim3.2\times{10}^{46}~{\rm erg}~{\rm Mpc}^{-3}~{\rm yr}^{-1}~\epsilon_{\rm cr,-1}L_{j,45}\rho_{\rm GC,-5}$, where $\rho_{\rm GC}$ is the density of GCs and $\epsilon_{\rm cr}$ is the CR energy fraction.  As shown in Fig.~6 of \cite{hil84}, they accelerate protons up to the maximum energy of $\varepsilon_p^{\rm max}\sim10\mbox{--}100$~EeV, overcoming various energy losses.  Then, CRs leaving AGN produce $pp$ neutrinos in large scale structures~\cite{gc}.  
 
In addition, during cosmological structure formation, large scale structures generate powerful IGSs on Mpc scales~\cite{igs}.  Strong shocks are expected around the virial radius $r_{\rm vir}\approx2.6~{\rm Mpc}~M_{15}^{1/3}$ (at $z=0$)~\cite{voi05}.  The accretion luminosity is $L_{\rm ac}\approx(\Omega_b/\Omega_m)GM\dot{M}/r_{\rm vir}\simeq0.9\times{10}^{46}~{\rm erg}~{\rm s}^{-1}~M_{15}^{5/3}$~\cite{igs}.  Taking $\rho_{\rm GC}\sim{10}^{-5}~{\rm Mpc}^{-3}$~\cite{GCcomment1}, the CR energy budget is $Q_{\rm cr}\sim1.0\times{10}^{47}~{\rm erg}~{\rm Mpc}^{-3}~{\rm yr}^{-1}~\epsilon_{\rm cr,-1}L_{\rm ac,45.5}\rho_{\rm GC,-5}$.  Using the typical shock radius $r_{\rm sh}\sim r_{\rm vir}$, shock velocity $V_s\sim{10}^{8.5}~{\rm cm}~{\rm s}^{-1}$ and magnetic field $B\sim0.1\mbox{--}1~\mu{\rm G}$~\cite{gcmag}, we have $\varepsilon_p^{\rm max}\approx(3/20)(V_s/c)eBr_{\rm sh}\sim1.2~{\rm EeV}~B_{-6.5}V_{s,8.5}M_{15}^{1/3}$~\cite{gai90} that can exceed 100~PeV.  

While CRs are injected by multiple AGN and/or IGSs for $t_{\rm inj}\sim$~a few~Gyr, the confined CRs produce neutrinos with hard spectra (even after $t_{\rm dyn}\approx r_{\rm sh}/V_s$ for an IGS).  For 100~PeV protons to be confined in GCs, the coherence length of $l_{\rm coh}\gtrsim0.34~{\rm kpc}~B_{-6.5}^{-1}\varepsilon_{p,17}$ is needed.  Assuming the Kolmogorov turbulence with $l_{\rm coh}\sim10\mbox{--}100$~kpc~\cite{gcmag}, we have the CR diffusion time, $t_{\rm diff}\approx(r_{\rm vir}^2/6D)\simeq1.6~{\rm Gyr}~\varepsilon_{p,17}^{-1/3}B_{-6.5}^{1/3}{(l_{\rm coh}/30~{\rm kpc})}^{-2/3}M_{15}^{2/3}$, which gives $\varepsilon_p^b\approx51~{\rm PeV}~B_{-6.5}{(l_{\rm coh}/30~{\rm kpc})}^{-2}M_{15}^2{(t_{\rm inj}/2~{\rm Gyr})}^{-3}$ from $t_{\rm diff}=t_{\rm inj}$.  The confinement of CRs with $\lesssim\varepsilon_p^b\sim100$~PeV can lead to hard spectra at $\lesssim\varepsilon_\nu^b\sim0.04\varepsilon_p^b\sim2$~PeV, while CRs with $\gtrsim\varepsilon_p^b$ escape into extracluster space, making neutrino spectra steeper at $\gtrsim\varepsilon_\nu^b$.  

Using typical intracluster densities $\bar{n}\sim{10}^{-4}~{\rm cm}^{-3}$~\cite{mb13,voi05}, with a possible enhancement factor $g\sim1\mbox{--}3$~\cite{mb13,GCcomment2}, we get $f_{pp}\simeq0.76\times{10}^{-2}~g\bar{n}_{-4}(t_{\rm int}/2~{\rm Gyr})$.  Then, we achieve $E_{\nu}^2\Phi_{\nu_i}\sim{10}^{-9}\mbox{--}{10}^{-8}~{\rm GeV}~{\rm cm}^{-2}~{\rm s}^{-1}~{\rm sr}^{-1}$, which can explain the INB flux~\cite{GCcomment3}.  
A neutrino break naturally arises from $t_{\rm diff}=t_{\rm inj}$.  Or, it may come from a broken power-law CR injection spectrum~\cite{GCcomment4,ks06} that has been suggested to explain CRs above 100~PeV~\cite{mur+08,ks06}. 

\subsection{Star-forming galaxies}
SFGs contain many supernova (SN) remnants that are promising CR accelerators.  Their CR budget is $Q_{\rm cr}\sim8.5\times{10}^{45}~{\rm erg}~{\rm Mpc}^{-3}~{\rm yr}^{-1}~\epsilon_{\rm cr,-1}\varrho_{\rm SFR,-2}$~\cite{lac13}.  The star-formation rate is $\varrho_{\rm SFR}\sim{10}^{-2}~M_\odot~{\rm Mpc}^{-3}~{\rm yr}^{-1}$ for main-sequence galaxies (MSGs) and $\varrho_{\rm SFR}\sim{10}^{-3}~M_\odot~{\rm Mpc}^{-3}~{\rm yr}^{-1}$ for SBGs~\cite{sfg}.  
At the Sedov radius $R_{\rm Sed}$, the proton maximum energy is $\varepsilon_p^{\rm max}\approx(3/20)(V_{\rm ej}/c)eBR_{\rm Sed}\simeq3.1~{\rm PeV}~B_{-3.5}{\mathcal E}_{\rm ej,51}^{1/3}V_{\rm ej,9}^{1/3}n^{-1/3}$, where ${\mathcal E}_{\rm ej}$ and $V_{\rm ej}$ are the ejecta energy and velocity.  SN shocks or their aggregation can achieve the {\it knee} energy when $B$ is high enough~\citep[e.g.,][]{hil84,stellarmag,sbgphysics}.   
The Galactic CR spectrum is dominated by heavy nuclei above the {\it knee}, so SFGs cannot explain the INB at $\gtrsim0.1$~PeV unless CRs are accelerated to higher energies in other galaxies.  But higher values $B\sim1\mbox{--}30$~mG indicated in SBGs~\cite{sbgsnrmag}, potentially give $\varepsilon_p^{\rm max}\sim100$~PeV.  Also, $\varepsilon_p^{\rm max}\gtrsim100$~PeV is expected for powerful supernovae (SNe) including hypernovae and trans-relativistic SNe~\cite{peculiarsne}.  Their fraction is typically a few percent of all SNe, but we note that they could be more common at higher redshifts and may contribute to the INB.  

Nearby SBGs like M82 and NGC 253 have a column density of $\Sigma_g \sim 0.1~{\rm g}~{\rm cm}^{-2}$ and a scale height of $h\sim50$~pc~\cite{sbgphysics}, while high-redshift starbursts in submillimeter galaxies have $\Sigma_g\sim1~{\rm g}~{\rm cm}^{-2}$ and $h\sim500$~pc~\cite{tac+06}, implying $\bar{n}\approx\Sigma_g/(2hm_p)\sim200~{\rm cm}^{-3}$.  High-redshift MSGs have $\Sigma_g\sim 0.1~{\rm g}~{\rm cm}^{-2}$ and $h \sim 1$~kpc~\cite{dad+10}, implying $\bar{n}\sim10~{\rm cm}^{-3}$.  
At low energies, CRs are confined in the starburst-driven wind (with its velocity $V_w$) and  advection governs escape, $t_{\rm esc}\approx t_{\rm adv}\approx h/V_w\simeq3.1~{\rm Myr}~(h/{\rm kpc})V_{w,7.5}^{-1}$.  Comparing with the pionic loss time $t_{pp}\approx2.7~{\rm Myr}~\Sigma_{g,-1}^{-1}~(h/{\rm kpc})$ gives $f_{pp} \approx1.1~\Sigma_{g,-1}V_{w,7.5}^{-1}(t_{\rm esc}/t_{\rm adv})$.  Therefore, CRs are significantly depleted by meson production during their advection~\cite{sbg,sbgphysics}.  
At higher energies, the diffusive escape becomes important~\cite{abr+12}.  
The confinement of 100~PeV protons requires the critical energy of $\varepsilon_c=eBl_{\rm coh}>100~{\rm PeV}$, leading to $l_{\rm coh}\gtrsim0.34~{\rm pc}~B_{-3.5}^{-1}\varepsilon_{p,17}$.  The diffusion coefficient at $\varepsilon_c$ is $D_c=(1/3)l_{\rm coh}c$, below which $D=D_c{(\varepsilon_p/\varepsilon_c)}^{\delta}$ (for $\delta\sim0\mbox{--}1$).    
Then, we have limits of $t_{\rm diff}\lesssim7.2~{\rm Myr}~B_{-3.5}^{-1}~(h /{\rm kpc})^2$ at 100~PeV and $D_0\gtrsim2.3\times{10}^{25}~{\rm cm}^2~{\rm s}^{-1}$ for $D=D_0{(\varepsilon_p/{\rm GeV})}^{1/3}$ in the Kolmogorov turbulence.  
The diffusion time is $t_{\rm diff}\approx(h^2/4D)\simeq1.6~{\rm Myr}~D_{0,26}^{-1}\varepsilon_{p,17}^{-1/3}{(h/\rm kpc)}^2$, giving $\varepsilon_p^b\approx21~{\rm PeV}~D_{0,26}^{-3}\Sigma_{g,-1}^3{(h/{\rm kpc})}^3$ (for $t_{pp}<t_{\rm adv}$) or $\varepsilon_p^b\approx15~{\rm PeV}~D_{0,26}^{-3}V_{w,7.5}^{3}{(h/{\rm kpc})}^3$ (for $t_{\rm adv}<t_{pp}$).    

If proton calorimetry largely holds~\cite{SFGcomment}, MSGs and SBGs may have $E_{\nu}^2 \Phi_{\nu_i}\sim{10}^{-9}\mbox{--}{10}^{-7}~{\rm GeV}~{\rm cm}^{-2}~{\rm s}^{-1}~{\rm sr}^{-1}$, sufficient for the INB flux~\cite{sbg}.  A break could come from $t_{\rm diff}=t_{pp}$ or $t_{\rm diff}=t_{\rm adv}$.  But we may simply expect a PeV cutoff due to $\varepsilon_\nu^{\rm cut}\sim0.04\varepsilon_p^{\rm max}$ for $\varepsilon_p^{\rm max}\sim100$~PeV (e.g., by hypernovae), where the locally observed CRs above $\sim100$~PeV would have different origins.  

\section{Summary and Discussion}
A crucial step towards revealing the origin of the IceCube signal is the discrimination between $pp$ and $p\gamma$ scenarios.  For $pp$ scenarios, combing the new IceCube and recent {\it Fermi} data leads to strong {\it upper} limits on $\Gamma$ and {\it lower} limits on the diffuse IGB contribution.  The results are largely independent of source models, redshift evolution and the existence of a multi-PeV neutrino break/cutoff.  They are the first strong constraints with the {\it measured} neutrino and $\gamma$-ray fluxes.  Further multimessenger studies in the near future can test the $pp$ scenarios by (a) determining $\Gamma$ by sub-PeV neutrino observations with IceCube, (b) improving our knowledge of the sub-TeV diffuse IGB, and (c) observing a number of the bright individual sources that should have hard spectra, by TeV $\gamma$-ray observations especially with CTA.  Also, IceCube may detect nearby GCs via stacking~\cite{mb13}, giving another test of the IGS scenario, while it seems difficult to see individual SFGs~\cite{sbgphysics}.  

We considered the origin of a possible break/cutoff, which is favored by the present data since $pp$ scenarios require $\Gamma\lesssim2.1\mbox{--}2.2$.  If it is real, it may provide clues to sources of observed CRs.  Neutrino sources are not necessarily related to such sources due to the low maximum energy, severe CR depletion and intervening magnetic fields.  But, as suggested in \cite{mur+08,ks06}, some models for observed CRs can have soft spectra of escaping CRs at $\gtrsim100$~PeV and hard neutrino spectra below PeV.   

Our results are useful for constructing specific source models.  For example, if the INB is explained by hypernovae in SFGs, contributions from normal SNe should not violate the IGB.  If $pp$ scenarios are ruled out by (a)-(c) in near future, $p\gamma$ scenarios like GRB and AGN models will be supported~\citep[c.f.][]{afterkyoto}.  
AGN jet models~\cite{agnjet} have difficulty since target photons in the infrared-to-optical range typically lead to $\sim10\mbox{--}1000$~PeV neutrinos, whereas  AGN core models~\cite{agncore} might work.  GRB inner jet models predict a neutrino break at $\sim1\mbox{--}100$~PeV due to the strong meson and muon cooling~\cite{grb} and potentially give contributions of $E_\nu^2\Phi_{\nu_i}\lesssim{10}^{-8}~{\rm GeV}~{\rm cm}^{-2}~{\rm s}^{-1}~{\rm sr}^{-1}$~\cite{grb}, but constraints from stacking searches by IceCube suggest $E_\nu^2\Phi_{\nu_i}\lesssim{10}^{-9}~{\rm GeV}~{\rm cm}^{-2}~{\rm s}^{-1}~{\rm sr}^{-1}$~\cite{grblim} so other possibilities, e.g., low-power GRBs~\cite{llgrb}, may be favored.  
Multimessenger tests for these sources, though much more model dependent, should be studied.  


\medskip
\begin{acknowledgments}
We thank Kfir Blum, Doron Kushnir, Francis Halzen, especially John Beacom, Todd Thompson, and Nathan Whitehorn for very encouraging comments and discussions.  This work is supported by NASA through Hubble Fellowship Grant No. 51310.01 awarded by the STScI, which is operated by the Association of Universities for Research in Astronomy, Inc., for NASA, under Contract No. NAS 5-26555 (K. M.), the U.S. National Science Foundation (NSF) under Grants No. OPP-0236449 and No. PHY-0236449 (M. A.), and a Jansky Fellowship from National Radio Astronomy Observatory that is operated by Associated Universities, Inc., under cooperative agreement with NSF (B. C. L.).
\end{acknowledgments}



\begin{thebibliography}{99}
\bibitem{GCcomment1}
The density of GCs with $\gtrsim{10}^{15}~M_\odot$ is $\sim{10}^{-6}~{\rm Mpc}^{-3}$, but is $\sim{10}^{-5}~{\rm Mpc}^{-3}$ for GCs above a few$\times{10}^{14}~M_\odot$~\cite{jen+01}.
\bibitem{GCcomment2}
High CR pressure in the GC core is constrained~\cite{radiolim}.
\bibitem{GCcomment3}
Predictions by \cite{mur+08} can be enhanced by a factor by including lower-mass structures with $\lesssim{10}^{15}~M_\odot$.
\bibitem{GCcomment4}
For instance, the CR spectrum may come from the superposition of hard spectra with $s$ and different maximum energies.  For $V_s\propto M^{1/3}$, constant $B$ and $d\rho_{\rm GC}/dM\propto M^{-\alpha}$ (for $\alpha\geq8/3$), we expect $s_2=s+3\alpha/2-4$.
\bibitem{SFGcomment}
Observations of M82 and NGC 253 indicate that only $20\mbox{--}40$\% of the CR power is lost to meson production~\cite{sbgphysics}.
\end{thebibliography}
\end{document}